\documentclass[12pt]{article}
\usepackage{amsfonts}

\newcommand{\END}{\end{document}}

\begin{document}

\begin{center}
\smallskip {\large FUNDAMENTALS OF QUANTUM}\

{\large MUTUAL\ ENTROPY AND\ CAPACITY}

\bigskip

\bigskip

Masanori Ohya

\bigskip

\medskip

Department of Information Sciences

Science University of Tokyo

Noda City, Chiba 278-8510, Japan

\smallskip
\end{center}

\section{Introduction}

The study of mutual entropy (information) and capacity in classical system
was extensively done after Shannon by several authors like Kolmogorov \cite
{Kol} and Gelfand \cite{GY}. In quantum systems, there have been several
definitions of the mutual entropy for classical input and quantum output
\cite{BS,Hol,Ing,Lev}. In 1983, the author defined \cite{O2} the fully
quantum mechanical mutual entropy by means of the relative entropy of
Umegaki \cite{Ume}, and he extended it \cite{O4} to general quantum systems
by the relative entropy of Araki \cite{Ara} and Uhlmann \cite{Uhl}. When the
author introduced the quantum mutual entropy, he did not indicate that it
contains other definitions of the mutual entropy including classical one, so
that there exist several misunderstandings for the use of the mutual entropy
(information) to compute the capacity of quantum channels. Therefore in this
note we point out that our quantum mutual entropy generalizes others and
where the misuse occurs.

\smallskip

\section{Mutual Entropy}

The quantum mutual entropy was introduced in \cite{O2} for a quantum input
and quantum output, namely, for a purely quantum channel, and it was
generalized for a general quantum system described by C*-algebraic
terminology \cite{O4}. We here review the mutual entropy in usual quantum
system described by a Hilbert space.

Let $\mathcal{H}$ be a Hilbert space for an input space,
$\bf{B}\mathcal{(H)}$ be the set of all bounded linear
operators on $\mathcal{H}$ and $\mathcal{\
S(H)}${\scriptsize \ }be the set of all density operators on
$\mathcal{H}.$ An output space is described by another Hilbert space $%
\stackrel{\sim }{\mathcal{H}}$ , but often $\mathcal{H=}\stackrel{\sim }{%
\mathcal{H}}$. A channel from the input system to the output system is a
mapping $\Lambda $* from $\mathcal{S(H)}$ to $\mathcal{S(\stackrel{\sim }{%
\mathcal{H}})}$ \cite{O1}. A channel $\Lambda $* is said to be completely
positive if the dual map $\Lambda $ satisfies the following condition: $%
\Sigma _{k,j=1}^{n}$ $A_{k}^{*}\Lambda (B_{k}^{*}B_{j})A_{j}\geq 0$ for any $%
n\in ${\bf{N}} and any $A_{j}\in B(\mathcal{H}),B_{j}\in
B(\stackrel{\sim  }{\mathcal{H}})$. This condition is not
strong at all because almost all physical transformations
satisfy it \cite{IKO,O4}.

An input state $\rho $ $\in \mathcal{S(H)}$ is sent to the output system
through a channel $\Lambda $*, so that the output state is written as $%
\stackrel{\sim }{\rho }\equiv \Lambda ^{*}\rho .$ Then it is important to
ask how much information of $\rho $ is correctly sent to the output state $%
\Lambda ^{*}\rho .$ This amount of information transmitted from input to
output is expressed by the mutual entropy.

In order to define the mutual entropy, we first mention the entropy of a
quantum state introduced by von Neumann \cite{Neu}. For a state $\rho ,$
there exists a unique spectral decomposition
\begin{equation}
\renewcommand{\theequation}{2.1}
\rho =\Sigma _{k}\lambda _{k}P_{k},
\end{equation}
where $\lambda _{k}$ is an eigenvalue of $\rho $ and $P_{k}$ is the
associated projection for each $\lambda _{k}$. The projection $P_{k}$ is not
one-dimensional when $\lambda _{k}$ is degenerated, so that the spectral
decomposition can be further decomposed into one-dimensional projections.
Such a decomposition is called a Schatten decomposition \cite{Sch}, namely,

\begin{equation}
\renewcommand{\theequation}{2.2}
\rho =\Sigma _{k}\lambda _{k}E_{k},
\end{equation}
where $E_{k}$ is the one-dimensional projection associated with $\lambda
_{k} $ and the degenerated eigenvalue $\lambda _{k}$ repeats dim$P_{k}$
times; for instance, if the eigenvalue $\lambda _{1}$has the degeneracy 3,
then $\lambda _{1}=\lambda _{2}=\lambda _{3}<\lambda _{4}$. This Schatten
decomposition is not unique unless every eigenvalue is non-degenerated. Then
the entropy (von Neumann entropy) $S\left( \rho \right) $ of a state $\rho $
is defined by

\begin{equation}
\renewcommand{\theequation}{2.3}
S\left( \rho \right) =-tr\rho \log \rho ,
\end{equation}
which equals to the Shannon entropy of the probability distribution $\left\{
\lambda _{k}\right\} $ :

\begin{equation}
\renewcommand{\theequation}{2.4}
S\left( \rho \right) =-\sum_{k}\lambda _{k}\log \lambda _{k}.
\end{equation}

The quantum mutual entropy was introduced on the basis of the above von
Neumann entropy for purely quantum communication processes. The mutual
entropy depends on an input state $\rho $ and a channel $\Lambda ^{*}$, so
it is denoted by $I\left( \rho ;\Lambda ^{*}\right) $, which should satisfy
the following conditions:

(1) The quantum mutual entropy is well-matched to the von Neumann entropy.
Furthermore, if a channel is trivial, i.e., $\Lambda ^{*}=$ identity map,
then the mutual entropy equals to the von Neumann entropy: $I\left( \rho
;id\right) $ = $S\left( \rho \right) $.

(2) When the system is classical, the quantum mutual entropy reduces to
classical one.

(3) Shannon's fundamental inequality \cite{Sha} 0$\leq $ $I\left( \rho
;\Lambda ^{*}\right) \leq S\left( \rho \right) $ is held.

Before mentioning the quantum mutual entropy, we briefly review the
classical mutual entropy \cite{Bil}. Let $\left( \Omega ,\mathcal{F}\right) $
, $\left( \overline{\Omega },\overline{\mathcal{F}}\right) $be an input and
output measurable spaces, respectively, and $P\left( \Omega \right) ,$ $%
P\left( \overline{\Omega }\right) $ are the corresponding sets of all
probability measures (states) on $\Omega $ and $\overline{\Omega }$ ,
respectively. A channel $\Lambda ^{*}$ is a mapping from $P\left( \Omega
\right) $ to $P\left( \overline{\Omega }\right) $ and its dual $\Lambda $ is
a map from the set $B\left( \Omega \right) $ of all Baire measurable
functions on $\Omega $ to $B\left( \overline{\Omega }\right) .$ For an input
state $\mu \in P\left( \Omega \right) ,$ the output state $\overline{\mu }=$
$\Lambda ^{*}\mu $ and the joint state (probability measure) $\Phi $ is
given by

\begin{equation}
\renewcommand{\theequation}{2.5}
\Phi \left( Q\times \;\;\overline{Q}\right)
=\int_{\overline{Q}}\Lambda
\left( 1_{Q}\right) d\mu ,\;\;Q\in
\mathcal{F},\;\;\overline{Q}\in
\overline{\mathcal{F}},
\end{equation}
where 1$_{Q}$ is the characteristic function on $\Omega :$ 1$_{Q}\left(
\omega \right) =\left\{
\begin{array}{ll}
1 & \left( \omega \in Q\right) \\
0 & \left( \omega \notin Q\right)
\end{array}
\right. .$ The classical entropy, relative entropy and mutual entropy are
defined as follows:

\begin{equation}
\renewcommand{\theequation}{2.6}
S\left( \mu \right) =\sup \left\{ -\sum_{k=1}^{n}\mu \left( A_{k}\right)
\log \mu \left( A_{k}\right) ;\left\{ A_{k}\right\} \in \mathcal{P}\left(
\Omega \right) \right\} ,
\end{equation}

\begin{equation}
\renewcommand{\theequation}{2.7}
S\left( \mu ,\nu \right) =\sup \left\{ \sum_{k=1}^{n}\mu \left( A_{k}\right)
\log \frac{\mu \left( A_{k}\right) }{\nu \left( A_{k}\right) };\left\{
A_{k}\right\} \in \mathcal{P}\left( \Omega \right) \right\} ,
\end{equation}

\begin{equation}
\renewcommand{\theequation}{2.8}
I\left( \mu ;\Lambda ^{*}\right) =S\left( \Phi ,\mu \otimes \Lambda ^{*}\mu
\right) ,
\end{equation}
where $\mathcal{P}\left( \Omega \right) $ is the set of all finite
partitions on $\Omega ,$ that is, $\left\{ A_{k}\right\} \in \mathcal{P}
\left( \Omega \right) $ iff $A_{k}$ $\in \mathcal{F}$ with $A_{k}\cap
A_{j}=\emptyset $ $\left( k\neq j\right) $and $\cup _{k=1}^{n}A_{k}=$ $%
\Omega .$

The quantum mutual entropy is defined as follows: In order to define the
quantum mutual entropy, we need the quantum relative entropy and the joint
state (it is called ''compound state'' in the sequel) describing the
correlation between an input state $\rho $ and the output state $\Lambda
^{*}\rho $ through a channel $\Lambda ^{*}$. A finite partition of $\Omega $
in classical case corresponds to an orthogonal decomposition $\left\{
E_{k}\right\} $ of the identity operator I of $\mathcal{H}$ in quantum case
because the set of all orthogonal projections is considered to make an event
system in a quantum system. It is known \cite{OP} that the following
equality holds

\[
\sup \left\{ -\sum_{k}tr\rho E_{k}\log tr\rho E_{k};\left\{ E_{k}\right\}
\right\} =-tr\rho \log \rho ,
\]
and the supremum is attained when $\left\{ E_{k}\right\} $ is a Schatten
decomposition of $\rho .$ Therefore the Schatten decomposition is used to
define the compound state and the quantum mutual entropy.

The compound state $\sigma _{E}$ (corresponding to joint state in CS) of $%
\rho $ and $\Lambda ^{*}\rho $ was introduced in \cite{O2,O3}, which is
given by

\begin{equation}
\renewcommand{\theequation}{2.9}
\sigma _{E}=\sum_{k}\lambda _{k}E_{k}\otimes \Lambda ^{*}E_{k},
\end{equation}
where $E$ stands for a Schatten decomposition $\left\{ E_{k}\right\} $ of $%
\rho ,$ so that the compound state depends on how we decompose the state $%
\rho $ into basic states (elementary events), in other words, how to see the
input state.

The relative entropy for two states $\rho $ and $\sigma $ is defined by
Umegaki \cite{Ume} and Lindblad \cite{Lin}, which is written as

\begin{equation}
\renewcommand{\theequation}{2.10}
S\left( \rho ,\sigma \right) =\left\{
\begin{array}{ll}
tr\rho \left( \log \rho -\log \sigma \right) & \left(
\hbox{when }\overline{ran\rho }\subset
\overline{ran\sigma }\right)
\\
\infty & \left( \hbox{otherwise}\right)
\end{array}
\right.
\end{equation}

Then we can define the mutual entropy by means of the compound state and the
relative entropy \cite{O2}, that is,

\begin{equation}
\renewcommand{\theequation}{2.11}
I\left( \rho ;\Lambda ^{*}\right) =\sup \left\{ S\left( \sigma _{E},\rho
\otimes \Lambda ^{*}\rho \right) ;E=\left\{ E_{k}\right\} \right\} ,
\end{equation}
where the supremum is taken over all Schatten decompositions because this
decomposition is not unique generally. Some computations reduce it to the
following form:

\begin{equation}
\renewcommand{\theequation}{2.12}
I\left( \rho ;\Lambda ^{*}\right) =\sup \left\{ \sum_{k}\lambda _{k}S\left(
\Lambda ^{*}E_{k},\Lambda ^{*}\rho \right) ;E=\left\{ E_{k}\right\} \right\}
.
\end{equation}
This mutual entropy satisfies all conditions (1)$\sim $(3) mentioned above.

When the input system is classical, an input state $\rho $ is given by a
probability distribution or a probability measure, in either case, the
Schatten decomposition of $\rho $ is unique, namely, for the case of
probability distribution ; $\rho =\left\{ \lambda _{k}\right\} ,$

\begin{equation}
\renewcommand{\theequation}{2.13}
\rho =\sum_{k}\lambda _{k}\delta _{k},
\end{equation}
where $\delta _{k}$ is the delta measure, that is,

\begin{equation}
\renewcommand{\theequation}{2.14}
\delta _{k}\left( j\right) =\delta _{k,j}=\{_{0(k\neq j)}^{1(k=j)},\forall j.
\end{equation}
Therefore for any channel $\Lambda ^{*},$ the mutual entropy becomes

\begin{equation}
\renewcommand{\theequation}{2.15}
I\left( \rho ;\Lambda ^{*}\right) =\sum_{k}\lambda _{k}S\left( \Lambda
^{*}\delta _{k},\Lambda ^{*}\rho \right) ,
\end{equation}
which equals to the following usual expression of Shannon when the minus is
well-defined:

\begin{equation}
\renewcommand{\theequation}{2.16}
I\left( \rho ;\Lambda ^{*}\right) =S\left( \Lambda ^{*}\rho \right)
-\sum_{k}\lambda _{k}S\left( \Lambda ^{*}\delta _{k}\right) .
\end{equation}
The above equality has been taken as the definition of the mutual entropy
for a classical-quantum channel \cite{B,BS,Hol,Ing,Lev}.

\smallskip Note that the definition (2.12) of the mutual entropy is written
as

\[
I\left( \rho ;\Lambda ^{*}\right) =\sup \left\{ \sum_{k}\lambda _{k}S\left(
\Lambda ^{*}\rho _{k},\Lambda ^{*}\rho \right) ;\rho =\sum_{k}\lambda
_{k}\rho _{k}\in F_{o}\left( \rho \right) \right\} ,
\]
where $F_{o}\left( \rho \right) $ is the set of all orthogonal finite
decompositions of $\rho .$ Here $\rho _{k}$ is orthogonal to $\rho _{j}$
(denoted by $\rho _{k}\perp \rho _{j})$ means that the range of $\rho _{k}$
is orthogonal to that of $\rho _{j.}$ The equality is easily proved as
follows: Put

\[
I_{f}\left( \rho ;\Lambda ^{*}\right) =\sup \left\{ \sum_{k}\lambda
_{k}S\left( \Lambda ^{*}\rho _{k},\Lambda ^{*}\rho \right) ;\rho
=\sum_{k}\lambda _{k}\rho _{k}\in F_{o}\left( \rho \right) \right\} .
\]
The inequality $I\left( \rho ;\Lambda ^{*}\right) \leq I_{f}\left( \rho
;\Lambda ^{*}\right) $ is obvious$.$ Let us prove the converse. Each $\rho
_{k}$ in an orthogonal decomposition of $\rho $ is further decomposed into
one dimensional projections; $\rho _{k}=\sum_{j}\mu _{j}^{\left( k\right)
}E_{j}^{\left( k\right) },$ a Schatten decomposition of $\rho _{k}$. From
the following equalities of the relative entropy \cite{Ara,OP}$:$ (1) $%
S\left( a\rho ,b\sigma \right) =$ $aS\left( \rho ,\sigma \right) -a\log
\frac{b}{a},$ for any positive number $a,$ $b;$ (2) $\rho _{1}\perp \rho
_{2} $ $\Longrightarrow S\left( \rho _{1}+\rho _{2},\sigma \right) =S\left(
\rho _{1},\sigma \right) +S\left( \rho _{2},\sigma \right) ,$ we have

$
\begin{array}{l}
\sum_{k}\lambda _{k}S\left( \Lambda ^{*}\rho _{k},\Lambda ^{*}\rho \right)
=\sum_{k,j}\lambda _{k}\mu _{j}^{\left( k\right) }S\left( \Lambda
^{*}E_{j}^{\left( k\right) },\Lambda ^{*}\rho \right) +\sum_{k,j}\lambda
_{k}\mu _{j}^{\left( k\right) }\log \mu _{j}^{\left( k\right) } \\
\leq \sum_{k,j}\lambda _{k}\mu _{j}^{\left( k\right) }S\left( \Lambda
^{*}E_{j}^{\left( k\right) },\Lambda ^{*}\rho \right) ,
\end{array}
$ which implies the converse inequality $I\left( \rho ;\Lambda ^{*}\right)
\geq I_{f}\left( \rho ;\Lambda ^{*}\right) $ because\\
$\sum_{k,j}\lambda _{k}\mu _{j}^{\left( k\right)
}E_{j}^{\left( k\right) }$ is a Schatten decomposition of
$\rho .$ Thus $I\left( \rho ;\Lambda ^{*}\right) =I_{f}\left(
\rho ;\Lambda ^{*}\right) .$

More general formulation of the mutual entropy for general quantum systems
was done \cite{O4,IKO} in C*dynamical system by using Araki's or Uhlmann's
relative entropy \cite{Ara,Uhl,OP}. This general mutual entropy contains all
other cases including measure theoretic definition of Gelfand and Yaglom
\cite{GY}.

\smallskip The mutual entropy is a measure for not only information
transmission but also description of state change, so that this quantity can
be applied to several topics \cite{AOS,Aka,MO1,Mur-O2,O4,O5,O10,OW}.

\smallskip

\section{Communication Processes}

We discuss communication processes in this section\cite
{Bil,Ing,OP}. Let $A=\{a_{1,}a_{2,}\cdot \cdot ,a_{n}\}$ be a set of certain
alphabets and $\Omega $ be the infinite direct product of $A:$ $\Omega
=A^{Z}\equiv \Pi _{-\infty }^{\infty }A$ calling a message space. In order
to send a information written by an element of this message space to a
receiver, we often need to transfer the message into a proper form for a
communication channel. This change of a message is called a coding.
Precisely, a coding is a measurable one to one map $\xi $ from $\Omega $ to
a proper space $X$ . For instance, we have the following codings: (1) When a
message is expressed by binary symbol 0 and 1, such a coding is a map from $%
\Omega $ to $\left\{ 0,1\right\} ^{N}.$ (2) A message expressed by 0,1
sequence in (1) is represented by an electric signal. (3) Instead of an
electric signal, we use optical signal. Coding is a combination of several
maps like the above (1) and (2), (3). One of main targets of the coding
theory is to find the most efficient coding and also decoding for
information transmission.

Let $\left( \Omega ,\mathcal{F}_{\Omega },P(\Omega )\right) $ be an input
probability space and $X$ be the coded input space. This space $X$ may be a
classical object or a quantum object. For instance, $X$ is a Hilbert space $%
\mathcal{H}$ of a quantum system, then the coded input system is described
by $\left( B(\mathcal{H)},\mathcal{S}(\mathcal{H)}\right) $of Sec.2.

An output system is similarly described as the input system: The coded
output space is denoted by $\stackrel{\sim }{X}$ and the decoded output
space is $\stackrel{\sim }{\Omega }$ made by another alphabets. An
transmission (map) from $X$ to $\stackrel{\sim }{X}$ is described by a
channel reflecting all properties of a physical device, which is denoted by $%
\gamma $ here. With a decoding $\stackrel{\sim }{\xi },$ the whole
information transmission process is written as
\begin{equation}
\renewcommand{\theequation}{3.1}
\Omega \stackrel{\xi }{\longrightarrow }X\stackrel{\gamma }{\longrightarrow }%
\stackrel{\sim }{X}\stackrel{\stackrel{\sim }{\xi }}{\longrightarrow }%
\stackrel{\sim }{\Omega }.
\end{equation}
That is, a message $\omega \in \Omega $ is coded to $\xi \left( \omega
\right) $ and it is sent to the output system through a channel $\gamma $,
then the output coded message becomes $\gamma \circ \xi \left( \omega
\right) $ and it is decoded to $\stackrel{\sim }{\xi }\circ \gamma \circ \xi
\left( \omega \right) $ at a receiver.

This transmission process is mathematically set as follows: M messages are
sent to a receiver and the $k$th message $\omega ^{\left( k\right) }$ occurs
with the probability $\lambda _{k}.$ Then the occurrence probability of each
message in the sequence $\left( \omega ^{\left( 1\right) },\omega ^{\left(
2\right) },\cdot \cdot \cdot ,\omega ^{\left( M\right) }\right) $of M
messages is denoted by $\rho =\left\{ \lambda _{k}\right\} ,$ which is a
state in a classical system. If $\xi $ is a classical coding, then $\xi
\left( \omega \right) $ is a classical object such as an electric pulse. If $%
\xi $ is a quantum coding, then $\xi \left( \omega \right) $ is a quantum
object (state) such as a coherent state. Here we consider such a quantum
coding, that is, $\xi \left( \omega ^{\left( k\right) }\right) $ is a
quantum state, and we denote $\xi \left( \omega ^{\left( k\right) }\right) $
by $\sigma _{k}.$ Thus the coded state for the sequence $\left( \omega
^{\left( 1\right) },\omega ^{\left( 2\right) },\cdot \cdot \cdot ,\omega
^{\left( M\right) }\right) $ is written as

\begin{equation}
\renewcommand{\theequation}{3.2}
\sigma =\sum_{k}\lambda _{k}\sigma _{k}.
\end{equation}
This state is transmitted through a channel $\gamma .$ This channel is
expressed by a completely positive mapping $\Gamma ^{*},$ in the sense of
Sec.1, from the state space of $X$ to that of $\stackrel{\sim }{X}$ , hence
the output coded quantum state $\stackrel{\sim }{\sigma }$ is $\Gamma
^{*}\sigma .$ Since the information transmission process can be understood
as a process of state (probability) change, when $\Omega $ and $\stackrel{
\sim }{\Omega }$ are classical and $X$ and $\stackrel{\sim }{X}$ are
quantum, the process (3.1) is written as

\begin{equation}
\renewcommand{\theequation}{3.3}
P\left( \Omega \right) \stackrel{\Xi ^{*}}{\longrightarrow }\mathcal{S}
\left( \mathcal{H}\right) \stackrel{\Gamma ^{*}}{\longrightarrow }\mathcal{S(%
}\stackrel{\sim }{\mathcal{H}})\stackrel{\stackrel{\sim }{\Xi }^{*}}{
\longrightarrow }P(\stackrel{\sim }{\Omega }),
\end{equation}
where $\Xi ^{*}$ $($resp.$\stackrel{\sim }{\Xi }^{*})$ is the channel
corresponding to the coding $\xi $ (resp.$\stackrel{\sim }{\xi }$ ) and $%
\mathcal{S}\left( \mathcal{H}\right) $ (resp.$\mathcal{S(}\stackrel{\sim }{%
\mathcal{H}})$ $)$ is the set of all density operators (states) on $\mathcal{%
\ H}$ (resp.$\stackrel{\sim }{\mathcal{H}}$ $)$.

We have to be care to study the objects in the above transmission process
(3.1) or (3.3). Namely, we have to make clear which object is going to
study. For instance, if we want to know the information capacity of a
quantum channel $\gamma (=\Gamma ^{*}),$ then we have to take $X$ so as to
describe a quantum system like a Hilbert space and we need to start the
study from a quantum state in quantum space $X\ $not from a classical state
associated to a message. If we like to know the capacity of the whole
process including a coding and a decoding, which means the capacity of a
channel $\stackrel{\sim }{\xi }\circ \gamma \circ \xi (=\stackrel{\sim }{\Xi
}^{*}\circ \ \Gamma ^{*}\circ \Xi ^{*})$, then we have to start from a
classical state$.$ In any case, when we concern the capacity of channel, we
have only to take the supremum of the mutual entropy $I\left( \rho ;\Lambda
^{*}\right) $ over a quantum or classical state $\rho $ in a proper set
determined by what we like to study with a channel $\Lambda ^{*}.$ We
explain this more precisely in the next section.

\smallskip

\section{Channel Capacity}

We discuss two types of channel capacity in communication processes, namely,
the capacity of a quantum channel $\Gamma ^{*}$ and that of a classical
(classical-quantum-classical) channel $\stackrel{\sim }{\Xi }^{*}\circ \
\Gamma ^{*}\circ \Xi ^{*}.$

(1) {\it{Capacity of quantum channel:}} The capacity of a
quantum channel is the ability of information transmission
of the channel itself, so that it does not depend on how to
code a message being treated as a classical object and we
have to start from an arbitrary quantum state and find the
supremum of the mutual entropy. One often makes a mistake
in this point. For example, one starts from the coding of a
message and compute the supremum of the mutual entropy
and he says that the supremum is the capacity of a quantum
channel, which is not correct. Even when his coding is a
quantum coding and he sends the coded message to a receiver
through a quantum channel, if he starts from a classical
state, then his capacity is not the capacity of the quantum
channel itself. In his case, usual Shannon's theory is applied
because he can easily compute the conditional distribution
by a usual (classical) way. His supremum is the capacity of a
classical-quantum-classical channel, and it is in the second
category discussed below.

The capacity of a quantum channel $\Gamma ^{*}$ is defined as follows: Let $%
\mathcal{S}_{0}(\subset $ $\mathcal{S(H))}$ be the set of all states
prepared for expression of information. Then the capacity of the channel $%
\Gamma ^{*}$ with respect to $\mathcal{S}_{0}$ is defined by

\begin{equation}
\renewcommand{\theequation}{4.1}
C^{\mathcal{S}_{0}}\left( \Gamma ^{*}\right) =\sup \{I\left( \rho ;\Gamma
^{*}\right) ;\rho \in \mathcal{S}_{0}\}.
\end{equation}
Here $I\left( \rho ;\Gamma ^{*}\right) $ is the mutual entropy given in
(2.11) or (2.12) with $\Lambda ^{*}=\Gamma ^{*}.$ When $\mathcal{S}_{0}=%
\mathcal{S(H)}$ , $C^{\mathcal{S}(\mathcal{H)}}\left( \Gamma ^{*}\right) $
is denoted by $C\left( \Gamma ^{*}\right) $ for simplicity. In \cite
{OPW1,Mur-O1}, we also considered the pseudo-quantum capacity $C_{p}\left(
\Gamma ^{*}\right) $ defined by (4.1) with the pseudo-mutual entropy $%
I_{p}\left( \rho ;\Gamma ^{*}\right) $ where the supremum is taken over all
finite decompositions instead of all orthogonal pure decompositions:
\begin{equation}
\renewcommand{\theequation}{4.2}
I_{p}\left( \rho ;\Gamma ^{*}\right) =\sup \left\{ \sum_{k}\lambda
_{k}S\left( \Gamma ^{*}\rho _{k},\Gamma ^{*}\rho \right) ;\rho
=\sum_{k}\lambda _{k}\rho _{k},\hbox{ finite
decomposition}\right\} .
\end{equation}
However the pseudo-mutual entropy is not well-matched to the conditions
explained in Sec.2, and it is difficult to be computed numerically \cite
{OPW2}. From the monotonicity of the mutual entropy \cite{OP}, we have

\[
0\leq C^{\mathcal{S}_{0}}\left( \Gamma ^{*}\right) \leq C_{p}^{\mathcal{S}
_{0}}\left( \Gamma ^{*}\right) \leq \sup \left\{ S(\rho );\rho \in \mathcal{S%
}_{0}\right\} .
\]

(2) {\it{Capacity of classical-quantum-classical channel:}}
The capacity of C-Q-C channel $\stackrel{\sim }{\Xi
}^{*}\circ \
\Gamma ^{*}\circ \Xi ^{*} $ is the capacity of the
information transmission process starting from the coding
of messages, therefore it can be considered as the capacity
including a coding (and a decoding). As is discussed in Sec.3,
an input state $\rho $ is the probability distribution $\left\{
\lambda _{k}\right\} $ of messages, and its Schatten
decomposition is unique as (2.13), so the mutual entropy is
written by (2.15):

\begin{equation}
\renewcommand{\theequation}{4.3}
I\left( \rho ;\stackrel{\sim }{\Xi }^{*}\circ \ \Gamma ^{*}\circ \Xi
^{*}\right) =\sum_{k}\lambda _{k}S\left( \stackrel{\sim }{\Xi }^{*}\circ \
\Gamma ^{*}\circ \Xi ^{*}\delta _{k},\stackrel{\sim }{\Xi }^{*}\circ \
\Gamma ^{*}\circ \Xi ^{*}\rho \right) .
\end{equation}
If the coding $\Xi ^{*}$ is a quantum coding, then $\Xi ^{*}\delta _{k}$ is
expressed by a quantum state. Let denote the coded quantum state by $\sigma
_{k}$ and put $\sigma =\Xi ^{*}\rho =\sum_{k}\lambda _{k}\sigma _{k}.$ Then
the above mutual entropy is written as

\begin{equation}
\renewcommand{\theequation}{4.4}
I\left( \rho ;\stackrel{\sim }{\Xi }^{*}\circ \ \Gamma ^{*}\circ \Xi
^{*}\right) =\sum_{k}\lambda _{k}S\left( \stackrel{\sim }{\Xi }^{*}\circ \
\Gamma ^{*}\sigma _{k},\stackrel{\sim }{\Xi }^{*}\circ \ \Gamma ^{*}\sigma
\right) .
\end{equation}
This is the expression of the mutual entropy of the whole information
transmission process starting from a coding of classical messages. Hence the
capacity of C-Q-C channel is

\begin{equation}
\renewcommand{\theequation}{4.5}
C^{P_{0}}\left( \stackrel{\sim }{\Xi }^{*}\circ \ \Gamma ^{*}\circ \Xi
^{*}\right) =\sup \{I\left( \rho ;\stackrel{\sim }{\Xi }^{*}\circ \ \Gamma
^{*}\circ \Xi ^{*}\right) ;\rho \in P_{0}\},
\end{equation}
where $P_{0}(\subset P(\Omega ))$ is the set of all probability
distributions prepared for input (a-priori) states (distributions or
probability measures). Moreover the capacity for coding is found by taking
the supremum of the mutual entropy (4.4) over all probability distributions
and all codings $\Xi ^{*}$:

\begin{equation}
\renewcommand{\theequation}{4.6}
C_{c}^{P_{0}}\left( \stackrel{\sim }{\Xi }^{*}\circ \ \Gamma ^{*}\right)
=\sup \{I\left( \rho ;\stackrel{\sim }{\Xi }^{*}\circ \ \Gamma ^{*}\circ \Xi
^{*}\right) ;\rho \in P_{0},\Xi ^{*}\}.
\end{equation}
The last capacity is for both coding and decoding and it is given by

\begin{equation}
\renewcommand{\theequation}{4.7}
C_{cd}^{P_{0}}\left( \ \Gamma ^{*}\right) =\sup \{I\left( \rho ;\stackrel{
\sim }{\Xi }^{*}\circ \ \Gamma ^{*}\circ \Xi ^{*}\right) ;\rho \in P_{0},\Xi
^{*},\stackrel{\sim }{\Xi }^{*}\}.
\end{equation}
These capacities $C_{c}^{P_{0}},$ $C_{cd}^{P_{0}}$ do not measure the
ability of the quantum channel $\Gamma ^{*}$ itself, but measure the ability
of $\Gamma ^{*}$ through the coding and decoding.

Remark that $\sum_{k}\lambda _{k}S(\Gamma ^{*}\sigma _{k})$ is finite, then
(4.4) becomes

\begin{equation}
\renewcommand{\theequation}{4.8}
I\left( \rho ;\stackrel{\sim }{\Xi }^{*}\circ \ \Gamma ^{*}\circ \Xi
^{*}\right) =S(\stackrel{\sim }{\Xi }^{*}\circ \Gamma ^{*}\sigma
)-\sum_{k}\lambda _{k}S(\stackrel{\sim }{\Xi }^{*}\circ \Gamma ^{*}\sigma
_{k}).
\end{equation}
Further, if $\rho $ is a probability measure having a density function $%
f(\lambda )$ and each $\lambda $ corresponds to a quantum coded state $%
\sigma (\lambda ),$ then $\sigma =\int f(\lambda )$ $\sigma (\lambda
)d\lambda $ and

\begin{equation}
\renewcommand{\theequation}{4.9}
I\left( \rho ;\stackrel{\sim }{\Xi }^{*}\circ \ \Gamma ^{*}\circ \Xi
^{*}\right) =S(\stackrel{\sim }{\Xi }^{*}\circ \Gamma ^{*}\sigma )-\int
f(\lambda )S(\stackrel{\sim }{\Xi }^{*}\circ \Gamma ^{*}\sigma (\lambda
))d\lambda ,
\end{equation}
which is less than

\[
S(\Gamma ^{*}\sigma )-\int f(\lambda )S(\Gamma ^{*}\sigma (\lambda
))d\lambda .
\]
The above bound is called Holevo bound, and it is computed in several cases%
\cite{OPW1,YO}.

The above three capacities $C^{P_{0}},$ $C_{c}^{P_{0}},$ $C_{cd}^{P_{0}}$
satisfy the following inequalities
\[
0\leq C^{P_{0}}\left( \stackrel{\sim }{\Xi }^{*}\circ \ \Gamma ^{*}\circ \Xi
^{*}\right) \leq C_{c}^{P_{0}}\left( \stackrel{\sim }{\Xi }^{*}\circ \
\Gamma ^{*}\right) \leq C_{cd}^{P_{0}}\left( \ \Gamma ^{*}\right) \leq \sup
\left\{ S(\rho );\rho \in P_{o}\right\}
\]
where $S(\rho )$ is not the von Neumann entropy but the Shannon entropy: -$%
\sum \lambda _{k}\log \lambda _{k}.$

The capacities (4.1), (4.6),(4.7) and (4.8) are generally different. Some
misunderstandings occur due to forgetting which channel is considered. That
is, we have to make clear what kind of the ability, the capacity of a
quantum channel itself or that of a classical-quantum(-classical ) channel
or that of a coding free, is considered.

\smallskip

\smallskip

\smallskip

\smallskip

\smallskip

\smallskip

\smallskip

\smallskip

\end{document}